\documentclass [12pt]{article}
\usepackage{amsmath}
\usepackage{amsfonts}
\usepackage{epsfig}
\begin{document}
\def\b{\bar}
\def\d{\partial}
\def\D{\Delta}
\def\cD{{\cal D}}
\def\cK{{\cal K}}
\def\f{\varphi}
\def\g{\gamma}
\def\G{\Gamma}
\def\l{\lambda}
\def\L{\Lambda}
\def\M{{\Cal M}}
\def\m{\mu}
\def\n{\nu}
\def\p{\psi}
\def\q{\b q}
\def\r{\rho}
\def\t{\tau}
\def\x{\phi}
\def\X{\~\xi}
\def\~{\widetilde}
\def\h{\eta}
\def\bZ{\bar Z}
\def\cY{\bar Y}
\def\bY3{\bar Y_{,3}}
\def\Y3{Y_{,3}}
\def\z{\zeta}
\def\Z{{\b\zeta}}
\def\Y{{\bar Y}}
\def\cZ{{\bar Z}}
\def\`{\dot}
\def\be{\begin{equation}}
\def\ee{\end{equation}}
\def\bea{\begin{eqnarray}}
\def\eea{\end{eqnarray}}
\def\half{\frac{1}{2}}
\def\fn{\footnote}
\def\bh{black hole \ }
\def\cL{{\cal L}}
\def\cH{{\cal H}}
\def\cF{{\cal F}}
\def\cP{{\cal P}}
\def\cM{{\cal M}}
\def\ik{ik}
\def\mn{{\mu\nu}}
\def\a{\alpha}

\title{Weakness of gravity as  illusion which hides true path to unification of  gravity with particle physics}

\author{Alexander Burinskii \\
NSI, Russian Academy of Sciences,\\ B. Tulskaya 52 Moscow 115191 Russia,
email: bur@ibrae.ac.ru}

\date{Mart 30, 2017 \fn{Essay received an Honorable Mention in the 2017 Essay Competition of the Gravity Research
Foundation.}} \maketitle

\begin{abstract}
Well known weakness of Gravity in particle physics is an illusion caused by underestimation of the role
   of spin in gravity. Relativistic rotation is inseparable from spin, which for elementary particles is extremely high and exceeds mass on 20-22 orders
   (in units $c=G=m=\hbar=1$). Such a huge spin  generates frame-dragging that distorts space much stronger than mass, and effective scale of gravitational interaction is shifted from Planck to Compton distances. We show that compatibility between gravity and quantum theory can be achieved without modifications of Einstein-Maxwell equations, by coupling to a supersymmetric Higgs model of  symmetry breaking and forming a nonperturbative super-bag solution, which generates a gravity-free Compton zone necessary for consistent work of quantum theory.
Super-bag is naturally upgraded to Wess-Zumino supersymmetric QED model, forming a bridge to perturbative formalism of conventional QED.
\end{abstract}

PACS: 12.10.-g, 12.60.Jv, 11.25.Sq, 11.27.+d
\maketitle

\newpage

As is known, Quantum theory and Gravity cannot be combined in
a unified theory. Gravity refuses pointlike, structureless  quantum particles, requiring extended field structure  for  right side
of Einstein equations, $G_\mn = 8\pi T_\mn .$

Revolutionary step for unification was made in superstring theory -- transition to extended stringlike objects, however,  \emph{``...Since
1974 superstring theory stopped to be considered as particle physics... ''}  and \emph{``... a
realistic model of elementary particles still appears to be a distant dream ... ''}, J. Schwarz \cite{Schw},

One of reason was the choice of  Planck scale as universal scale
for all unifications, including gravity.
 Attempt to bring gravitational scale close to the weak scale was considered in the braneworld scenario, where the weakness of the localized 4d gravity was explained by its ``leaks'' into higher-dimensional bulk. Braneworld mechanism allowed
to realize ideas of the superstring theory for any numbers of extra dimensions \cite{DvalBW}.

Alternative approach was related with solitons -- nonperturbative 4D solutions of the nonlinear field models, in particular solutions to low energy string theory \cite{Dabh,Sen,BurSen}.
This approach, being akin to Higgs mechanism of symmetry breaking, is matched with
nonperturbative approach to electroweak sector of Standard Model, where the most known are the
Nielsen-Olesen model of dual string based on the Landau-Ginzburg (LG) field model for superconducting media, and the famous MIT and SLAC bag models \cite{MIT,SLAC,Dash}  which are similar to solitons, but being soft, deformable and oscillating,
acquire many properties of the string models. The question on
consistency with gravity is not discussed usually for the solitonic models, as it is conventionally
assumed that gravity is weak and not essential at scale of electroweak interactions (see, for example, \cite{Malda,Baez})

We claim that assumption on weakness of gravity is an illusion, related with
 underestimation of the role of spin in gravity.  In relativistic theory spin is inseparable from rotation, and created by spin invariant effect of gravitational frame-dragging \cite{MTW} (supported by Probe B experiment), or Lense-Thirring effect in Kerr geometry \cite{Kerr}, distorts space along with mass.

 Spin of elementary particles is extremely high. In particular, for electron spin/mass ratio is about $10^{22} $ (in dimensionless units $G=c=\hbar=1 $), and its influence becomes so strong that conflict with quantum theory is shifted from Planck  to Compton scale.\fn{In the used `natural'  Planck's units the Planck's mass,  length and energy are $M_P =l_P =E_P = 1 ,$ and the energy equivalent to unit of spin $\hbar =1 $ is equal to Planck energy  $E_p =1 .$} Similar to Cosmology where giant masses turn gravity into a main force, GIANT SPIN of particles MAKES GRAVITY STRONG and crucial by its interplay with quantum theory.

 The spinning Kerr-Newman (KN)  solution \cite{Kerr,DKS} is of particular interest, since it has gyromagnetic ratio  $g=2 ,$
 corresponding to Dirac theory of electron \cite{Car},  and structure of the KN solution with such a huge spin sheds light on the reason of conflict between gravity and quantum theory and points out the way for its resolution.

Metric of KN solution in the Kerr-Schild form is \cite{DKS}
\be g_\mn =\eta_\mn + 2H k_\m k_\n , \label{KS}\ee where $ \eta_\mn $ is metric of auxiliary
Minkowski space $M^4 ,$ (signature $(- + + +)$), where scalar function
 \be H_{KN}=\frac {mr - e^2/2}{r^2+a^2 \cos ^2 \theta} ,\label{HKN} \ee is given in
oblate spheroidal coordinates $r$ and $\theta$, determined
  by transformations \cite{DKS},
  \be x+iy = (r + ia) \exp \{i\phi_K \} \sin \theta , \quad
z=r\cos \theta, \quad \rho =r-t . \label{coordKerr} \ee

The null field $ k_\m (x) ,$   ($ k_\m k^\m =0 $)  determines directions of dragging of space, Fig.1.

\noindent Lense-Thirring effect creates vortex of congruence $ k_\m (x) ,$  which for ultra-high spin, $a = J/m >> m ,$  becomes so strong that BH horizons disappear, and  $k^\m (x)$ focus on singular ring $r=0, \cos\theta =0 ,$  forming branch line of space into two sheets,
$ g_\mn^\pm =\eta_\mn + 2H k_\m^\pm k_\n^\pm ,$ defined by ingoing $k^-_\n  $ and outgoing congruence $ k^+_\n  .$

\begin{figure}[ht]
\centerline{\epsfig{figure=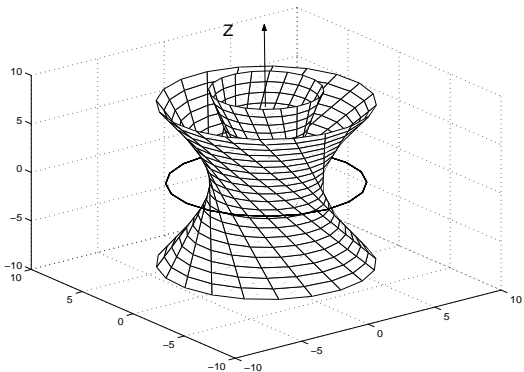,height=5cm,width=7cm}} \caption{The lightlike Kerr
congruence $k^\m$ determines space-dragging caused by mass and spin. Two sheets of Kerr metric
correspond to  $r<0 ,$ and  $r>0 $.}
\end{figure}

 Quantum theory requires flat space, at least in Compton zone, but electron spin $J=\hbar/2 $ exceeds mass $m$ about 22 orders, which  breaks space topologically, creating singular ring of Compton radius $ a = \hbar/2m .$
Singularity is signal to new physics. Usually, it is considered as signal to modify gravity.   We suggest alternative solution based on \emph{supersymmetry,} which  expels gravitational field  from  Compton zone of spinning particle, similar to expulsion of electromagnetic field from superconductor. Supersymmetric bag model, \cite{BurBag,BurBag1,Bur50}, realizes such expulsion of gravity and electromagnetic field, forming tree zones:

\textbf{(I)} -- flat quantum interior,

 \textbf{(E)} -- external zone with exact KN solution,

 \textbf{(R)} -- zone of transition from \textbf{(I)} to \textbf{(E)}.

\noindent For the giant values of spin, these demands become so restrictive that
structure of bag is determined almost unambiguously.

Surface $(R)$ is defined by the continuous transition of KN solution to Minkowski interior of the bag, (C. L\'opez \cite{Lop}).
According (\ref{KS}) and (\ref{HKN}), zone \textbf{(R)} corresponds to \be H_{KN}(r)=0 .\label{HKNzg} \ee
which gives
\be r = R = \frac {e^2}{2m} , \label{Hre}\ee and relations (\ref{HKNzg}) and (\ref{coordKerr}) show that bag takes
 form of a disk with thickness $ R $ and radius  $r_c = \sqrt {R^2 + a^2} ,$ Fig.2.

\begin{figure}[ht]
\centerline{\epsfig{figure=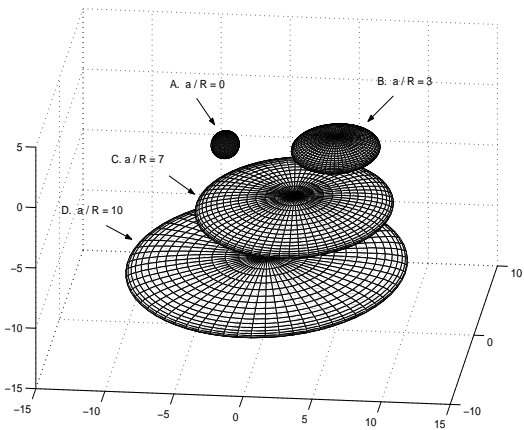,height=6cm,width=7cm}} \caption{Shape of disk for different $a=J/m$: (A) -  $a/R =0$, (B)-
$ a/R =3$; \ (C) - $ a/R =7$; and (D) - $ a/R =10$.} \label{fig2}
\end{figure}

To satisfy \textbf{(I),(E),(R)}, it is natural to use Higgs mechanism of symmetry breaking which is used in many nonperturbative
electroweak nodels, \cite{Solit}, and also in the MIT and SLAC bag models \cite{MIT,SLAC}.  The corresponding Lagrangian is also known as Landau-Ginzburg (LG) field model for superconducting phase transitions. The famous Nielsen-Olesen (NO) model for vortex string
in superconducting media, \cite{NO}, is based on LG Lagrangian
 \be
{\cal L}_{NO}= -\frac 14 F_\mn F^\mn - \frac 12 (\cD_\m \Phi)(\cD^\m \Phi)^* - V(|\Phi|),
\label{LNO}\ee where $ \cD_\m = \nabla_\m +ie A_\m $  are  $U(1)$ covariant derivatives,
$F_\mn = A_{\m,\n} - A_{\n,\m} $ the corresponding field strength, and potential $V$ has typical form \be V = \lambda ( \Phi ^\dag \Phi - \eta^2)^2 , \label{VNO} \ee
  where $\eta$ is condensate of Higgs field  $\Phi ,$  $\eta = <|\Phi|> $.

However, potential (\ref{VNO}) distorts external KN solution, placing Higgs field in  zone (E).
 It turns out that conditions \textbf{(I),(E),(R)} are satisfied  by supersymmetric LG model with three Higgs-like fields, \cite{Mor}
  $ (H, Z, \Sigma ) \equiv (\Phi_1, \Phi_2, \Phi_3) .$

Corresponding Lagrangian differs from  (\ref{LNO}) only
by summation over the fields  $\Phi_i , i=1,2,3,$ while the potential $V$
is changed very essentially, and  formed from a superpotential function $W(\Phi_i),$ \cite{WesBag}
  \be V(r)=\sum _i F_i F_i ^*, \quad  F_i = \d W /\d \Phi_i \equiv \d_i W  ,\label{VFi} \ee
where
   \be W(\Phi_i, \bar \Phi_i) = Z(\Sigma \bar
\Sigma -\eta^2) + (Z+ \m) H \bar H, \label{WLG}. \ee

\noindent The conditions  $F_i=\d_i W =0 $
determine $ two \ vacuum \ states, V=0 $ :
\begin{description}
  \item  [(I)] - internal: $ r<R-\delta, \ where \ Higgs \ field  |H| = \eta ,$
  \item [(E)] - external:  $ r>R +\delta , \  where \ Higgs \ field  H=0, $
\end{description}
 separated by $ zone \ of \ phase \ transition $ \textbf{(R)}, $ V > 0 ,$ in correspondence with the requirements\textbf{(I),(E),(R)}.
  \begin{figure}[ht]
\centerline{\epsfig{figure=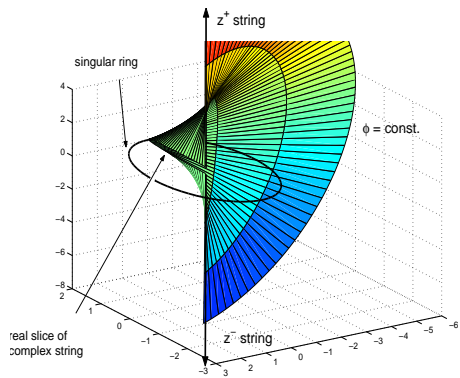,height=5cm,width=6cm}} \caption{Rotational dragging of  potential $A^\m$, forms Wilson loop along border of disk.}
\end{figure}

  Bag models with  potential (\ref{VNO}) form "cavity in superconductor", \cite{Dash}. Supersymmetric potential (\ref{VFi})-(\ref{WLG}) concentrates Higgs field in zone \textbf{(I)}, forming supersymmetric and superconducting vacuum state inside bag, where (\ref{LNO})  gives equation
\be \Box A_\m = J_\m =  e |H|^2 (\chi,_\m + e A_\m)
\label{Main} ,\ee showing that  current  $ J_\m$ vanishes inside the bag, and is concentrated (as usual, \cite{LL8}) in surface layer of superconducting disk.

Near boundary of disk $r = R = e^2/2m , \ \cos\theta =0 ,$  vector-potential $ A_\m $  is dragged  by  Kerr congruence (Fig.3), forming closed Wilson loop along singular ring.  \emph{It has remarkable consequence --  quantization of angular
 momentum,} \cite{BurBag1,Bur50,BurSol},
 $ J =n/2,  \quad n=1,2,3, ... $.

\bigskip

  Bag models take intermediate position between strings and solitons
\cite{Giles,JT,Tye}. Similar to solitons, they are nonperturbative solutions of the Higgs field model, but they have several specific features, in particular, flexibility  and ability to create string-like structures.  Under rotation,
 bags are deformed and take shape of stringy flux-tube joining the
quark-antiquark pair \cite{MIT,JT}, or toroidal string \cite{SLAC,Tye,GilTye}.

Spinning gravitational field sets shape of bag according \textbf{(R)},
and circular string is formed on the boundary of the disk, closely to  Kerr singular ring  (Fig.4A). So, the string is really formed by singular ring and regularized by bag boundary. The assumption, that Kerr singular ring is similar to NO model of dual string was done  long ago in \cite{Bur0,IvBur}, where it was noted that excitations of the KN solution create traveling waves along
 the Kerr ring. Later, it was obtained  in \cite{BurSen}  close connection of the Kerr singular ring
 with the Sen fundamental string solution to low energy string theory \cite{KerSen}, and other relations of Kerr geometry with string theory  \cite{BurStr}.
 String admits traveling waves, which  deform position
  of the bag boundary according \textbf{(R)}, creating a circulating lightlike node, where
   surface of the deformed bag touches the Kerr singular ring, creating a circulating lightlike singular pole, which can be associated with a confined quark, Fig.4B, and super-bag  turns into a single ``bag-string-quark'' system, analog of $D2-D1-D0$-brane  of  string--Mtheory.

\begin{figure}[ht]
\centerline{\epsfig{figure=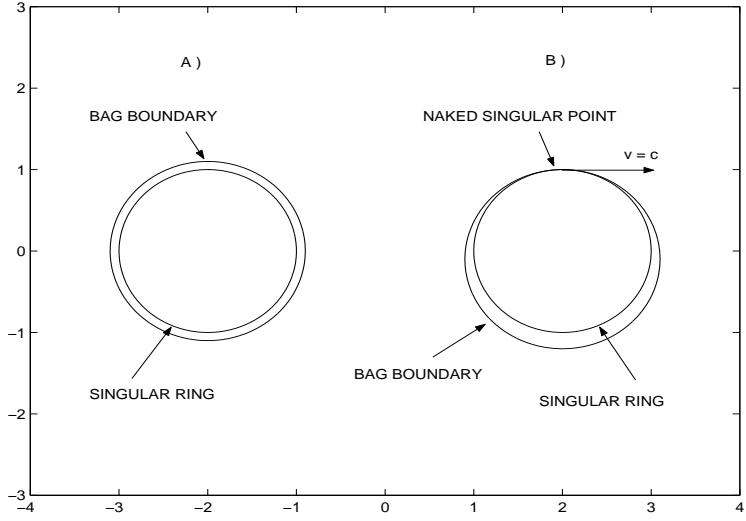,height=7cm,width=10cm}} \caption{Regularization of the KN
string.  Boundary of  bag fixes cut-off $R=r_e $ for the Kerr singular ring: A)  exact KN
solution, B)  KN solution excited by the lowest traveling mode creates singular
pole.} \label{fig.7}
\end{figure}

In turn, capture of quarks, is one more  special feature of the bag model, requiring consistent implementation of the Dirac equation \cite{MIT,SLAC,Tye,GilTye}. In KN geometry, it is defined according to famous Kerr Theorem \cite{DKS,PenRin} which defines the shear-free Kerr congruences in twistor terms, and gives two roots $Y^\pm $ for projective spinor coordinate \be Y =\phi_1/\phi_0 , \label{Y10}
\ee which is equivalent to two-component Weyl spinor $\phi_\alpha
,$ and defines the null direction as
   \be k_\m = \bar
\phi_{\dot\alpha} \sigma_\m^{\dot\alpha \alpha}\phi_\alpha . \ee

As it was shown in \cite{BurDirKN,BurBag}, two roots of the Kerr theorem $Y^\pm $ give us two Weyl spinors, $ \phi _\alpha $ and $
\bar\chi ^{\dot \alpha},$ consistent with ingoing and outgoing  KN solutions in  zone \textbf{(E)}.
Inside the bag, these solutions unite, forming the Dirac spinor
\be \Psi = \left(\begin{array}{cc}
 \phi _\alpha \\
\bar\chi ^{\dot \alpha}
\end{array} \right)\ee
which gets mass through Yukawa coupling to condensate of the Higgs
field.

\noindent Here we meet third  feature of the bag model -- emergence of the position-dependent mass term $m=G |\Phi| $, which is determined by spacetime distribution of the Higgs field.

 The Dirac equation,
 $\gamma^\m \d_\m \Psi = m \Psi ,$
 which is massive inside the bag,
turns out to be massless, and splits into two independent massless equations
\be
 \sigma ^\m _{\alpha \dot \alpha} i \d_\m
 \bar\chi ^{\dot \alpha}=0 , \quad
 \bar\sigma ^{\m \dot\alpha \alpha} i \d_\m
 \phi _{\alpha} = 0,
\label{Dir0} \ee
outside the bag,
corresponding to the left-handed and right-handed ``electron-type leptons'' of the
Glashow-Salam-Weinberg model \cite{GSWeinb}.

Finally, Super-Bag  can be naturally upgraded to Wess-Zumino SuperQED model, \cite{SuperBag},
revealing connections between the non-perturbative solutions of the supersymmetric LG model and the
conventional perturbative technics used in QED.

\bigskip

\textbf{Conclusion:}

\noindent Weakness of Gravity is delusion caused by the underestimation of huge impact of spin on space-time metric and topology. Disposal of this delusion opens a supersymmetric way to unify Gravity with particle physics.

\end{document}